\documentclass [twocolumn, aps, showpacs] {revtex4}
\usepackage{graphicx}
\usepackage{amsmath}
\usepackage{amssymb}
\usepackage{epsf}
\newlength{\upit}\upit=0.1truein

\newcommand{\ltappr}{{{\lower4pt\hbox{$<$} } \atop \widetilde{ \ \ \ }}}
\newlength{\bxwidth}\bxwidth=1.5 truein

\def\be{\begin{equation}}
\def\ee{\end{equation}}
\def\eqa{\begin{eqnarray}}
\def\eea{\end{eqnarray}}

\newlength{\figwidth}
\newlength{\shift}
\newcommand \bea {\begin{eqnarray} }

\begin{document}
\title {Solving Satisfiability Problems by the Ground-State Quantum Computer}

\author {Wenjin Mao}
\affiliation{ Department of Physics and Astronomy, Stony
Brook University, Stony Brook, NY 11794, U.S.A.}
\date{\today}
\begin{abstract}
A quantum algorithm is proposed to solve the Satisfiability problems by the ground-state quantum computer. The scale of the energy gap of the ground-state quantum computer is analyzed for the 3-bit Exact Cover problem. The time cost of this algorithm on the general SAT problems is discussed.
\end{abstract}
\maketitle

\section{Introduction}
A quantum computer has been expected to outperform its classical counterpart in some computation problems. For example, the well-known Shor's factoring algorithm\cite{Shor}  and Grover's algorithm\cite{Grover} accelerate exponentially and quadratically compared with the classical algorithms, respectively.
It is  a challenge to find whether a quantum computer outperforms on other classically intractable problems\cite{Farhi1, Hogg}, which cannot be solved classically in polynomial time of $N$, the number of the input bits.

Especially interesting are  the NP-complete problems\cite{NP-Complete}, which include thousands of problems, such as the Traveling Salesman problem\cite{EC} and the satisfiability (SAT) problems.   All NP-complete problems can be transformed into each other in polynomial steps. If one of the NP-complete problems can be solved in polynomial time by an algorithm even in the worst case, then all NP-complete problems can be solved in polynomial time. However, it is widely believed that such a classical algorithm doesn't exist. In this paper we will discuss quantum algorithm for solving SAT problems. A $K$-SAT problem deals with $N$ binary variables submitted to $M$ clauses with each clause $C_i$  involving $K$ bits, and the task is to find $N$-bit states satisfying all clauses.
 When $K>2$, $K$-SAT is NP-Complete, and some instances become
classically intractable when the parameter $\alpha=M/N$, as $M,\ N\rightarrow\infty$, approaches the threshold $\alpha_c(K)$\cite{3SAT, nature, SAT, NP}. 

Due to the properties of quantum mechanics, it's hard to design quantum algorithms directly from intuition. In the present paper, we will study the properties of the ground-state quantum computer(GSQC), and show that the special property of the GSQC naturally leads to algorithm for solving SAT problems.  Although we cannot determine whether or not this algorithm solves the NP-complete problems in polynomial time, we  try to shed light on the complexity of the NP-complete problems.

In the following sections, at first we introduce the idea of the ground-state quantum computer\cite{Mizel1, Mizel2, Mizel3} and its energy gap analysis\cite{ours}, then demonstrate the particular property of the GSQC, which provides a direct approach to solving SAT problems, and finally an example, an algorithm for solving the 3-bit Exact Cover problem, is given.

\section{ Ground-State Quantum Computer and its Energy Gap}

  A standard computer is characterized by a time-dependent state 
$
|\psi(t_i)\rangle=U_i|\psi(t_{i-1})\rangle,
$
where $t_i$ denotes the instance of  the $i$-th step, and $U_i$ represents for a  unitary transformation. For a GSQC, the time sequence is mimicked by  the spatial distribution of its ground-state wavefunction $|\psi_0\rangle$. 
As proposed by Mizel et.al.\cite{Mizel1}, the time evolution of a qubit may be represented by a column of  quantum dots with multiple  rows, and each row contains a pair of quantum dots. State $|0\rangle$ or $|1\rangle$ is represented by finding the electron in one of the two dots. It is important to notice that only one electron exists in a qubit. The energy gap, $\Delta$, between the first excited state and the ground state determines the scale of time cost.

\subsection{Hamiltonians of GSQC}
A GSQC is a  circuit of multiple interacting qubits, whose ground state is determined by the summation of the single qubit unitary transformation Hamiltonian $h^j(U_j)$, the two-qubit interacting Hamiltonian $h(CNOT)$, the boost Hamiltonian $h(B,\lambda)$ and the projection Hamiltonian $h(|\gamma\rangle,\lambda)$. 

The single qubit unitary transformation Hamiltonian has the form
\eqa
h^j(U_j)=\epsilon\left[ 
C^{\dagger}_{j-1}C_{j-1}+C^{\dagger}_{j}C_{j} -\left(C^{\dagger}_{j}U_jC_{j-1}+h.c.\right)\right],
\eea
where  $\epsilon$ defines the energy scale of all Hamiltonians, $C^{\dagger}_j=\left[c^{\dagger}_{j,0}\ c^{\dagger}_{j,1}\right]$, $c_{j,0}^{\dagger}$ is the electron creation operator on row $j$ at position $0$, and $U_j$ is a two dimension matrix representing the unitary transformation from row $j-1$ to row $j$. The boost  Hamiltonian is
\eqa
h^j(B,\lambda)&=&\epsilon\left[
C^{\dagger}_{j-1}C_{j-1}+\frac{1}{\lambda^2}C^{\dagger}_{j}C_{j}\right. \nonumber\\
& & \left. \ \ \ \ \ \ \ \ \ \ 
-\frac{1}{\lambda}\left(C^{\dagger}_{j}C_{j-1}+h.c.\right)\right],
\eea
which amplifies the wavefunction amplitude by the large value number $\lambda$ compared with  the  previous row at $|\psi_0\rangle$. The projection Hamiltonian is
\eqa
h^j\left(|\gamma\rangle,\lambda\right)&=&\epsilon\left[
c^{\dagger}_{j-1,\gamma}c_{j-1,\gamma}+\frac{1}{\lambda^2}c^{\dagger}_{j,\gamma}c_{j,\gamma} \right.
\nonumber\\
&& \ \ \ \ \ \ \ \ \ \ 
\left. -\frac{1}{\lambda}\left(c^{\dagger}_{j,\gamma}c_{j-1,\gamma}+h.c.\right)\right], 
\eea
where $|\gamma\rangle$ is the state to be projected to on row $j$ and to be amplified by $\lambda$ at $|\psi_0\rangle$. The interaction between qubit $\alpha$ and $\beta$ can be represented by $h(CNOT)$:
\eqa
&& \ h^j_{\alpha,\beta}(CNOT)
\nonumber\\
&=&\epsilon C^{\dagger}_{\alpha,j-1} C_{\alpha,j-1} C^{\dagger}_{\beta,j} C_{\beta,j}
+h^j_{\alpha}(I)C^{\dagger}_{\beta,j-1} C_{\beta,j-1}\nonumber\\
&&
+c^{\dagger}_{\alpha,j,0} c_{\alpha,j,0}h^j_{\beta}(I)
+c^{\dagger}_{\alpha,j,1} c_{\alpha,j,1}h^j_{\beta}(N). \label{hCNOT}
\eea
where for $c^{\dagger}_{a,b,\gamma}$, its subscription $a$ represents for qubit $a$, $b$ for the number of row, $\gamma$ for the state $|\gamma\rangle$. 
 With only $h^j(U_j)$ and $h^j_{\alpha,\beta}(CNOT)$, its ground state is\cite{Mizel2}:
\eqa
|\psi_0^j\rangle&=&\left[
1+c^{\dagger}_{\alpha,j,0}c_{\alpha,j-1,0}\left(1+C^{\dagger}_{\beta,j}C_{\beta,j-1}\right)\right.
\nonumber\\
&&\left. +c^{\dagger}_{\alpha,j,1}c_{\alpha,j-1,1}\left(1+C^{\dagger}_{\beta,j}NC_{\beta,j-1}\right)\right]\nonumber\\
&&\times \prod_{a\ne \alpha,\beta}\left(1+C^{\dagger}_{a,j}U_{a,j}C_{a,j-1}\right)|\psi^{j-1}\rangle.
\label{cnot}
\eea
All above mentioned  Hamiltonians are positive semidefinite, and are the same as those in \cite{Mizel1, Mizel2, Mizel3}. Only pairwise interaction is considered.

 The input states are determined by the boundary conditions applied upon the first rows of all qubits, which can be Hamiltonian $h^0=E(I+\sum_i a_i\sigma_i)$ with $\sigma_i$ being Pauli matrix and $\sum_ia_i^2=1$. For example, with $h^0=E(I+\sigma_z)$, $|\psi_0\rangle$ on the first row is $|1\rangle$; with $h^0=E(I-\sigma_x)$,  it is $ \left(|0\rangle+|1\rangle\right)$.  If $E$ is large enough, for example, at $E\ge 10 \epsilon$, the energy gap will saturate and become independent of the magnitude of $E$ \cite{ours}.

To implement an algorithm, on final row of each qubit a boost or a projection Hamiltonian is applied so that $|\psi_0\rangle$ concentrates on the position corresponding to the final instance in the standard paradigm,  hence measurement on the GSQC can read out the desired information with appreciable probability. With boost Hamiltonian or projection Hamiltonian on last rows, the ground-state wavefunction amplitude on those rows will be $\lambda$ of that on their neighboring rows.

By observing the expression Eq.(\ref{cnot}), it's easy to find that, for two interacting qubits, the ground-state wavefunction has the form\cite{ours}
\eqa
 &&\left( |\psi^{control}_{upstream}\rangle +|\psi^{control}_{downstream} \rangle \right) |\psi^{target}_{upstream}\rangle\nonumber\\
&&+ |\psi^{control}_{downstream}\rangle |\psi^{target}_{downstream}\rangle,
\label{wavefunction}
\eea
where each qubit is divided by the interacting Hamiltonian as two parts, and the part with boundary Hamiltonian $h^0$ is called as upstream, and the other part is called downstream. In this paper, we always use this definition when upstream or downstream is mentioned.

\subsection{Energy Gap of GSQC}
Now we briefly introduce how to find the scale of the energy gap of a GSQC. For details, please find in \cite{ours}.

With multiple interacting qubits, one needs to evaluate  on each qubit the parameter $1/x$, the overall amplitude of lowest excited state on top rows of this qubit before meeting  the first interacting Hamiltonian, assuming that on the top rows of this qubit the lowest energy excited state is orthonormal to $|\psi_0\rangle$ while states on all other qubits remain the same as the corresponding ground state with only magnitude changed. The energy gap\cite{ours} is given by the minimum parameter $1/x$ as
\eqa
\Delta\propto\epsilon(1/x)_{min}^2. 
\eea

The rule of estimating $1/x$ is as following\cite{ours}:
With each qubit ended with either a projection or a boost Hamiltonian containing the same (for simplicity) amplifying factor $\lambda\gg 1$, when estimating $1/x$  for a qubit, say qubit $A$, ($i$) at first $x$ is set to 1; ($ii$) the boost Hamiltonian,  $not$ the $projection$ Hamiltonian, on qubit $A$ itself increases $x$ by multiplication of $\lambda$; ($iii$) if qubit $A$ directly interacts with another qubit, say qubit $B$, by Hamiltonian $h_{AB}$, then we determine, excluding qubit $A$, on the qubit $B$ the ground-state wavefunction amplitude ratio of the upstream part (with respect to $h_{AB}$)  over its final row, $\frac{1}{x_B}$, contributions to $\frac{1}{x_B}$ are found one by one according to Eq.(\ref{wavefunction}): if the upstream part of qubit $B$ doesn't coexist with the states on final rows of any one qubit, except for qubit $A$, then $x_B$ should be multiplied by a $\lambda$; ($iv$) finally, the value of $1/x$ on qubit $A$ should be multiplied by $\frac{1}{x_B}$, or $\Pi_{i}\frac{1}{x_B^i}$ if more than one qubit directly interact with qubit $A$.

According to the above rule, the energy gap $\Delta$ of single qubit with length $n$ and ended with boost Hamiltonian $h(B,\lambda)$ scales as $\epsilon/\lambda^2$ as $\lambda\gg n$; when ended with projection Hamiltonian $h(|\gamma\rangle,\lambda)$, $\Delta$ is independent of $\lambda$. For two $n$-row qubits interacting by $h(CNOT)$, $\Delta\propto \epsilon/\lambda^4$ as $\lambda\gg n$ if both qubits ended with $h(B,\lambda)$ or one with $h(B,\lambda)$ and the other with $h(|\gamma\rangle,\lambda)$. Numerical calculations confirm these results.
The Fig.(1b) and Fig.(2) in \cite{ours} are two examples on how to apply the above rule on complicated circuits.

\begin{figure}
\begin{center}
\leavevmode
\hbox{\epsfxsize=6cm \epsffile{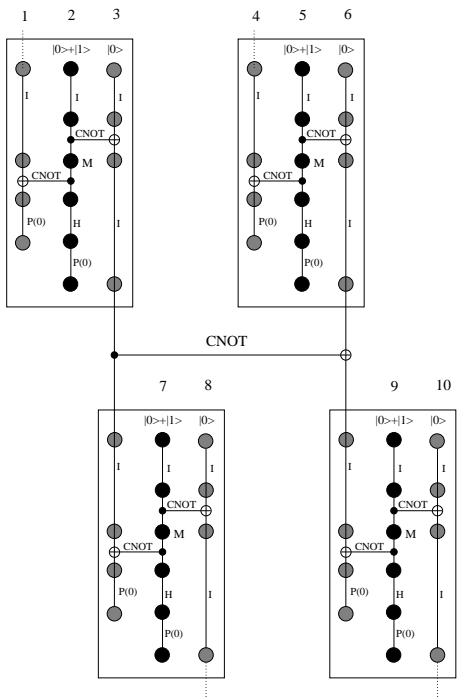}}
\end{center}
\caption{\small The same as Fig.(2) in \cite{ours}, this figure shows, in a complicated circuit, how the CNOT interacting qubits is modified by inserting teleportation boxes on each qubit's upstream and downstream part, so that the energy gap is only polynomially small.  Each dot represents a row of two quantum dots, label $I$ stands for identical transformation Hamiltonian $h(I)$, $H$ for Hadamard transformation Hamiltonian $h(H)$, and $P(0)$ for projection Hamiltonian $h(|0\rangle,\lambda)$. }
\label{teleport}
\end{figure}

Complicated GSQC circuit may have exponentially small energy gap, like the circuit in Fig.(1b) of \cite{ours}, and assembling the GSQC circuit directly following the algorithm for the standard paradigm, such as quantum Fourier transform, leads to exponentially small energy gap. In order to avoid such small gap, the teleportation boxes are  introduced on each qubit between two control Hamiltonians\cite{ours}. Fig.(\ref{teleport}) shows how the CNOT interacting qubits is modified by inserting teleportation boxes on each qubit's upstream and downstream part. The teleportation boxes make all qubits short (the longest qubit has length 8), on the other hand, for arbitrary GSQC circuit they make the energy gap only polynomially small $\Delta\propto \epsilon/\lambda^8$\cite{ours} if all boost and projection Hamiltonians have the same amplifying factor $\lambda$. To determine magnitude of $\lambda$, one only needs to count the total number of qubits in the circuit, say $L$, which is proportional to the number of control operation in an algorithm, then the probability of finding all electrons on final rows is $P\approx (1-C/\lambda^2)^{L}$ with $C$ being 8, the maximum length of qubit. In order to have appreciable $P$, we set $\lambda\approx L^{1/2}$, hence $\Delta\propto \epsilon/L^{4}$. The details can be found in \cite{ours}.

\subsection{Energy Gap When Projecting Small Fraction of a State} \label{fraction}

\begin{figure}
\begin{center}
\leavevmode
\hbox{\epsfxsize=6cm \epsffile{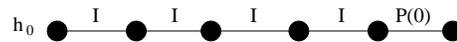}}
\end{center}
\caption{\small A six-row single qubit ended with the projection Hamiltonian $h(|0\rangle,\lambda)$. }
\label{Projection}
\end{figure}

\begin{figure}
\begin{center}
\leavevmode
\hbox{\epsfxsize=8cm \epsffile{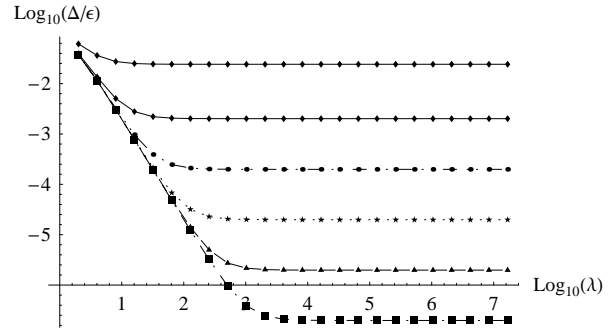}}
\end{center}
\caption{\small Energy gap $\Delta$ verse $\lambda$ with $h(|0\rangle,\lambda)$ applied on the last row of a 6-row single qubit, where $h^0=10\epsilon(I+\alpha\sigma_z-\sqrt{1-\alpha^2}\sigma_x)$. From top to bottom, lines correspond to $\alpha=0,\ 0.9,\ 0.99,\ 0.999,\ 0.9999,\ 0.99999$.}
\label{ProjEn}
\end{figure}

In the previous section the rule for finding scale of the energy gap is under the assumption that when a projection Hamiltonian $h(|\gamma\rangle,\lambda)$ is applied, $|a|/\sqrt{|a|^2+|b|^2}$ is appreciable for the ground state on row just before the projection Hamiltonian: 
\eqa
a|\gamma\rangle+b|\tilde{\gamma}\rangle,
\label{ab}
\eea 
where $|\gamma\rangle=|0\rangle$ ($|1\rangle$) and $|\tilde{\gamma}\rangle=|1\rangle$ ($|0\rangle$).  The ground-state wavefunction concentrates on the last row, hence the first excited state wavefunction cannot have appreciable weight there because otherwise $\langle \psi_1|\psi_0\rangle \ne 0$. When evaluate $1/x$ on a qubit, the projection Hamiltonian on the qubit itself doesn't contribute to $1/x$. For example, concerning a single qubit, as shown in Fig.(\ref{Projection}), with only identical transformations $h(I)$ and ended by $h(|0\rangle,\lambda)$, if $h^0=E(I-\sigma_x)$ so that $|\psi_0\rangle$ on the first row is $|0\rangle+|1\rangle$, then the energy gap $\Delta$ is almost independent of $\lambda$, as shown in the top line of Fig.(\ref{ProjEn}).

However, if in Eq.(\ref{ab}) $|a|/ \sqrt{|a|^2+|b|^2} \ll 1$, then $\Delta$ depends on $\lambda$ until $\lambda$ reaching $\sqrt{|a|^2+|b|^2} /|a|$. This is because when $\lambda< \sqrt{|a|^2+|b|^2} /|a|$, the ground-state wavefunction has little weight on the last row, and the first excited state concentrates there, hence $1/x$ is small, leading to small energy gap. When $\lambda> \sqrt{|a|^2+|b|^2} /|a|$, ground state wavefunction has large part on the last row, then just like the above situation, energy gap is not further affected by increasing $\lambda$.

To confirm the above analysis, we numerically calculate the energy gap of a 6-row single qubit ended with projection Hamiltonian, as shown in Fig.(\ref{Projection}).
 The boundary Hamiltonian is $h^0=10\epsilon(I+\alpha\sigma_z-\sqrt{1-\alpha^2}\sigma_x)$, all other Hamiltonians except for that at final row are $h^j(I)$ with $j=1,\ 2,\ 3,\ 4$, and on the final row there is a projection Hamiltonian $h(|0\rangle,\lambda)$. By tuning $\alpha$, we can determine what fraction of wavefunction is projected from the 5th row to the last row. At $\alpha=0,\ 0.9,\ 0.99,\ 0.999,\ 0.9999,\ 0.99999$, on the 5th row the ground state wavefunctions are $a|0\rangle+b|1\rangle$ with $a/b=1,\ 0.23,\ 0.071,\ 0.022,\ 0.0071,\ 0.0022.$ Fig.(\ref{ProjEn}) shows that the energy gap is  $\Delta\propto\epsilon/\lambda^2$ as $\lambda<|\sqrt{|a|^2+|b|^2}/a|$, and when $\lambda>|\sqrt{|a|^2+|b|^2}/a|$, $\Delta$ becomes independent on $\lambda$. The independent $\Delta$ is proportional to $\epsilon |a/\sqrt{|a|^2+|b|^2}|^2$.

 In order to make the ground-state wavefunction concentrate on the last row so that measurement corresponds to the desired state, $\lambda$ must be larger than $| \sqrt{|a|^2+|b|^2} /a|$, Thus the energy gap is determined by the fraction of state been projected. If $|a/ \sqrt{|a|^2+|b|^2} |$ is exponentially small, which may happen in certain case, then the energy gap is exponentially small. Fortunately, this doesn't happen to the GSQC implement of Quantum Fourier Transform, there all projection Hamiltonians are applied to teleportation circuit, and $|a/b|=1$. However, it plays a role in the algorithm presented in the following section.

For multiple interacting qubits, if $|a/ \sqrt{|a|^2+|b|^2} |\ll 1$ in Eq.(\ref{ab}), the rule of finding energy gap needs modification: 
With all qubits ended with either a projection or a boost Hamiltonian containing the same amplifying factor $\lambda\gg 1$, when estimating $1/x$  for any qubit, say qubit $A$, ($i$) at first $x$ is set to 1; ($ii$) the boost Hamiltonian,  $or$ the $projection$ Hamiltonian, on qubit $A$ itself increases $x$ by multiplication of $\lambda$ or Min$(\lambda,|\sqrt{|a|^2+|b|^2}/a|)$; ($iii$) if qubit $A$ directly interacts with another qubit, say qubit $B$ by Hamiltonian $h_{AB}$, then we determine, excluding qubit $A$, on the qubit $B$ the amplitude ratio of the upstream part (divided by $h_{AB}$)  over its final row, $1/x_B$, and contribution to $1/x_B$ from other qubits are found one by one according to Eq.(\ref{wavefunction}): if the upstream part of qubit $B$ doesn't coexist with the states on final rows of a qubit, except for qubit $A$, then $x_B$ should be multiplied by  $\lambda$ (ended with boost Hamiltonian) or $\lambda|a''/\sqrt{|a''|^2+|b''|^2}|$ (ended with projection Hamiltonian); ($iv$) finally, the value of $1/x$ on qubit $A$ should be multiplied by $1/x_B$ or $\Pi_i{1/x_B^i}$ if more than one qubit directly interact with qubit $A$.

It is easy to find that when $|b/a|$, $|b''/a''|\approx 1$ and $\lambda\gg 1$, we get the same result as the previous subsection.
After $1/x$'s on all qubits being evaluated, the minimum $1/x$ gives the energy gap scale as 
\eqa
\Delta\propto\epsilon(1/x)_{min}^2. \nonumber
\eea

\section{Quantum Algorithm by GSQC}

There are some interesting properties for the GSQC. Although it was shown\cite{Lloyd} that, concerning on time cost, a quantum computer composed of (time varying) local Hamiltonians is equivalent to standard circuit  quantum computer, GSQC provides some insights to design quantum algorithm for certain problems. For example, the projection Hamiltonian, which corresponds to measurement in standard paradigm, can amplify the probability at a particular state. Here we are not claiming that the GSQC is more powerful than standard quantum computer, however, the GSQC does provide a direct approach for certain problem, as shown below is the algorithm for the SAT problems.

\begin{figure}
\begin{center}
\leavevmode
\hbox{\epsfxsize=3cm \epsffile{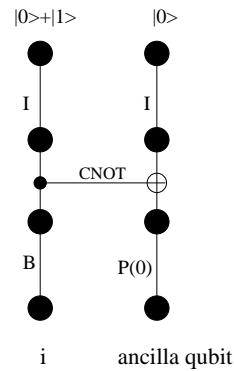}}
\end{center}
\caption{\small A filter for the clause $i=0$.}
\label{Two}
\end{figure}

At first we give the simplest example, considering that qubit $i$ $CNOT$ controls an ancilla qubit that is at the right side in  Fig.(\ref{Two}), and their boundary Hamiltonians make the ground state on their first rows are $|0\rangle+|1\rangle$ and $|0\rangle$, respectively. On last rows the ground state is $|0\rangle|0\rangle+|1\rangle|1\rangle$. If we apply a boost Hamiltonian on qubit $i$ and a projection Hamiltonian $h(|0\rangle,\lambda)$ on the ancilla qubit, then at the ground state the state on final rows becomes $|0\rangle|0\rangle$. The large value of $\lambda$ makes sure that there is large probability to find two electrons on the final rows of the two qubits at the ground state. So by choosing projected state on the ancilla qubit, we can have the selected state $|0\rangle$ on qubit $i$, and prevent the other state $|1\rangle$ from reaching its final row. If qubit $i$ entangles with other qubit, such as $|0\rangle|\alpha\rangle+|1\rangle|\beta\rangle$, the entanglement  of $|0\rangle|\alpha\rangle$ will not be affected. Thus we call circuit in Fig.(\ref{Two}) a filter for the clause $i=0$.

\begin{figure}
\begin{center}
\leavevmode
\hbox{\epsfxsize=3cm \epsffile{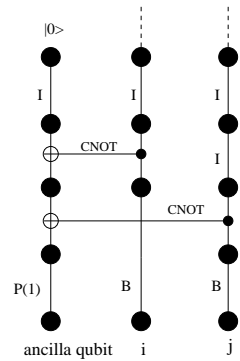}}
\end{center}
\caption{\small A filter for the clause $i+j=1$.}
\label{antiferro}
\end{figure}

Another example makes more sense. Lets consider a SAT problem with clauses, each of which involves two qubits, say qubit $i$ and $j$, and requires $i+j=1$. We can implement this clause by the GSQC circuit in Fig.(\ref{antiferro}). In this figure there are three qubits: qubit $i$, qubit $j$ and an ancilla qubit that is at the left side in the figure. It's easy to find that if on the first row $|i\rangle=\alpha_i|0\rangle+\beta_i|1\rangle,\ |j\rangle=\alpha_j|0\rangle+\beta_j|1\rangle$ and the ancilla qubits  at $|0\rangle$, then at the ground state on the final rows of the three qubits the state is $|i\rangle|j\rangle|ancilla\rangle=(\beta_i\alpha_j|1\rangle|0\rangle+\alpha_i\beta_j|0\rangle|1\rangle)|1\rangle$, which satisfies the clause. Thus circuit in Fig.(\ref{antiferro}) filters out states not satisfying this simple clause and lets through those satisfying states. It is important to note  that at the beginning if the satisfying states entangle with other qubits not showing in the figure, these entanglements keep untouched.



\begin{figure}
\begin{center}
\leavevmode
\hbox{\epsfxsize=6cm \epsffile{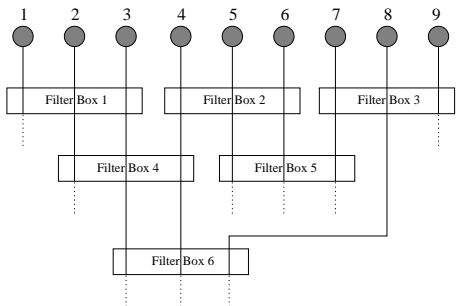}}
\end{center}
\caption{\small A GSQC circuit solving SAT problem with each clause involving several bits. Box labeled ``Filter Box" represents filter like Fig.(\ref{Two})(one-bit clause), Fig.(\ref{antiferro})(two-bit clause) or Fig.(\ref{filterbox})(three-bit clause).}
\label{Circuit}
\label{circuit}
\end{figure}

The property of GSQC brings up new quantum algorithm naturally. Here we present one to solve the SAT problems as shown in Fig.(\ref{Circuit}),
a GSQC circuit to solve  a 3-SAT problem with only 9 bits. It's easy to be extended to $N$-bit $K$-SAT problems. Each clause is implemented by a ``filter box", and the circuit inside each filter box makes sure that on rows immediately below it the ground state satisfies the  clause $C_i$, or we can say those unsatisfying states are filtered out. This can be realized by projection and boost Hamiltonians like in Fig.(\ref{Two}) and Fig.(\ref{antiferro}).

In Fig.(\ref{Circuit}), the initial state on the top rows of qubit from 1 to 9 is $(|0\rangle+|1\rangle)(|0\rangle+|1\rangle)...(|0\rangle+|1\rangle)$, which is enforced by the boundary Hamiltonians, $h^0=E(I-\sigma_x)$;
the clause involving qubit 1, 2 and 3 is implemented by filter box 1, the clause involving qubit 2, 3 and 4 implemented by filter box 4, the clause involving qubit 3, 4 and 8 implemented by filter box 6, etc.


When all constraints are implemented, at ground state the states measured on the final rows of the $N$ qubits should be superposition of all states satisfying all constraints. No  backtracking is needed.

Concerning energy gap, unlike the circuit for quantum Fourier transform, in which the energy gap is determined by the number of control operation\cite{ours}, the SAT problems is more complicated to evaluate because it might involve the situation to project a very small fraction of state as shown in section \ref{fraction}. For example, if one constructs a GSQC for the Grover's search problem with one condition to find a unique satisfying state from $2^N$ states, then he will find that there is an ancilla qubit containing such unnormalized state 
\eqa
|0\rangle|\text{satisfying}\rangle+\sum_{i=1}^{2^{N}-1}|1\rangle|\text{unsatisfying}^{(i)}\rangle
\label{fff}
\eea
before the projection Hamiltonian $h(|0\rangle,\lambda)$. In order to amplify the amplitude of the correct state on the final row, it requires $\lambda\ge 2^{N/2}$. Its energy gap is hence less than $2^{-N}$, which is consistent with the limit set by many other works\cite{Grover, Farhi2, Bennett}.


\section{Example: the 3-Bit Exact Cover Problem}

Up to now the filters, Fig.(\ref{Two}) and Fig.(\ref{antiferro}), we have given are trivial, and now we give an example on how to implement a filter for a serious problem.
We  focus  on the 3-bit Exact Cover problem\cite{EC},  an instance of SAT problem, which belongs to NP-complete. Following is the definition of the 3-bit Exact Cover problem:

{\textit{There are $N$ bits $z_1,\ z_2,\ ...,\ z_N$, each taking the value 0 or 1. With $O(N)$   clauses  applied to them, each clause is a constraint involving three bits: one bit has value 1 while the other two have value 0. The task is to determine the $N$-bit state satisfying all the clauses. }}

\subsection{GSQC Circuit for the 3-bit Exact Cover Problem}

\begin{figure}
\begin{center}
\leavevmode
\hbox{\epsfxsize=5cm \epsffile{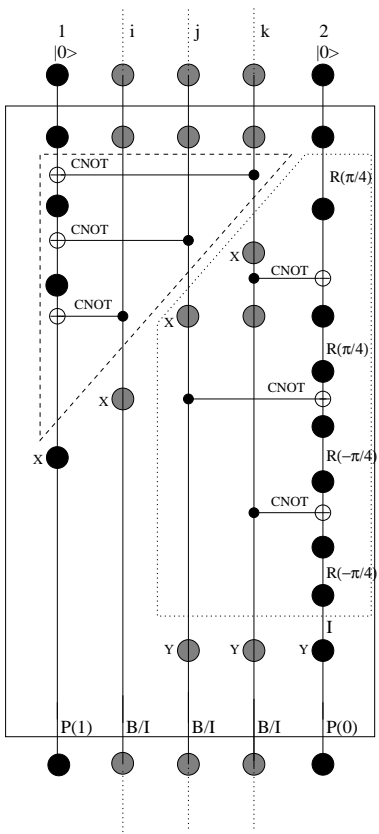}}
\end{center}
\caption{\small A filter for the clause $i+j+k=1$. The labels  on the lines stand for corresponding Hamiltonians: $I$ for $h(I)$, $CNOT$ for $h(CNOT)$, $P(1)$ for projection $h(|1\rangle,\lambda)$ et. al. At the final rows, $B/I$ represents boost Hamiltonian $h(B,\lambda)$ if there is no more clause to be applied to this qubit, otherwise, represents identical transformation Hamiltonian $h(I)$. There are teleportation boxes, not shown in figure, inserted on all qubits between two control Hamiltonians. Some dots marked by $X$ or $Y$ are for demonstration convenience in text.}
\label{filterbox}
\end{figure}

The algorithm is implemented by the circuit in Fig.(\ref{circuit}). Each filter box, in our algorithm, involves three qubits, say qubit $i,\ j$ and $ k$, which are represented by gray dot columns in Fig.(\ref{filterbox}). We add two ancilla qubits: qubit 1 and qubit 2, which are represented by dark dot columns.   Qubit $i,\ j$ and $k$ at the first row are in the state $(|1\rangle+|0\rangle)$ if they have not experienced any clause yet, and the two ancilla qubits are in the states $|\hat{0}\rangle$ and $|\tilde{0}\rangle$ on top rows by selecting proper boundary Hamiltonians, where $ |\hat{\gamma}\rangle$ corresponds to the state of ancilla qubit 1, and $ |\tilde{\gamma}\rangle$ to the state of ancilla qubit 2. 

Inside the dashed triangle of Fig.(\ref{filterbox}),
after the first $CNOT$, we obtain state  $|\hat{1}{\rangle}|1{\rangle} + |\hat{0}{\rangle}|0{\rangle}$;
after the second $CNOT$:  $ |\hat{1}{\rangle}|1{\rangle}|0{\rangle}+|\hat{0}{\rangle}|0{\rangle}|0{\rangle} +  |\hat{0}{\rangle}|1{\rangle}|1\rangle+|\hat{1}{\rangle}|0{\rangle}|1{\rangle}$;
after the third $CNOT$: 
\eqa
&& |\hat{1}\rangle \left(|1\rangle |0\rangle |0\rangle+|0\rangle |1\rangle |0\rangle+|0\rangle |0\rangle |1\rangle+|1\rangle |1\rangle |1\rangle\right)\nonumber\\
&+&|\hat{0}\rangle \left(|1\rangle |1\rangle |0\rangle+|0\rangle |1\rangle |1\rangle+|1\rangle |0\rangle |1\rangle+|0\rangle |0\rangle |0\rangle\right).\nonumber
\eea
Immediately below the triangle, if the system stays at the ground state, if electron in ancilla qubit 1 is measured to be on the row labeled by $X$ and at state $|\hat{1}\rangle$, and if the three electrons on qubit $i,\ j,\ k$ are all found on the rows labeled by $X$, then the three-qubit state satisfies the clause except for $ |1{\rangle}|1{\rangle}|1{\rangle}$.

The ancilla qubit $2$, starting at state $ |\tilde{0} {\rangle}$, experiences  $ CNOT$ gates controlled by qubits $j$ and $k$, and $R(\pm \pi/4)$ transformations, defined in \cite{Tof} as $R_y(\pm \pi/4)$, as shown within the dotted pentagon in Fig.(\ref{filterbox}). All those transformations happened  inside the dotted pentagon are equivalent to a Toffoli gate except for some unimportant phases\cite{Tof}: if both qubits $j$ and $k$ are in state $|1\rangle$, then the ancilla qubit $2$ reverses to state $|\tilde{1}\rangle$, otherwise, it remains at state $|\tilde{0}\rangle$. After this nearly Toffoli transformation, if at ground state electrons in qubit $j,\ k$ and ancilla qubit $2$ are found on rows labeled by $Y$, and if ancilla qubit 2 is at $|\tilde{0}\rangle$, then the three qubits will be at $|\tilde{0} \rangle ( |0 \rangle |0\rangle + |1\rangle |0\rangle + |0\rangle |1\rangle)$.
 Thus if at ground state all electrons are found on rows immediately below both the dashed triangle and the dotted pentagon, and if ancilla qubit 1 is at $|\hat{1}\rangle$ and ancilla qubit 2 at $|\tilde{0}\rangle$, then  the three qubits $i,\ j,\ k$  satisfy the clause:
\eqa
|\hat{1}\rangle|\tilde{0}\rangle\left(|1\rangle|0\rangle|0\rangle+|0\rangle|1\rangle|0\rangle+|0\rangle|0\rangle|1\rangle
\right).
\label{ff}
\eea

In order to make the satisfying states pass through the filter box with large probability, we add projection Hamiltonians and boost Hamiltonians as shown in the lower part of Fig.(\ref{filterbox}).
The projection Hamiltonians on final rows of the two ancilla qubits limit and amplify the amplitude of the states we prefer:   ancilla qubit 1 at $|\hat{1}\rangle$, and   ancilla qubit 2 at $|\tilde{0}\rangle$.
 If a qubit does not experience any more clause, it will end with a boost Hamiltonian, otherwise, its quantum state will be teleported to a new qubit through teleportation box, not shown in Fig.(\ref{filterbox}), and the new qubit experiences more clauses. Thus the projection Hamiltonians on two ancilla qubits and boost Hamiltonians on the three qubits make sure that the ground-state wavefunction concentrates on the final rows in Fig.(\ref{filterbox}) with state at Eq.(\ref{ff}).

 Noting that in the filter box all the three qubits $i,\ j,$ and $k$ always act as control qubits, thus the entanglements of these three qubits with other qubits not involved in this particular clause still keep the same. When adding a clause, the resulted states satisfying this clause will also satisfy all previous applied clauses. Thus unlike classical algorithm, no backtracking is needed.

\subsection{Energy Gap Without Projecting Small Fraction of State}
In this subsection, we assume applying each clause does decrease the number of satisfying state $gradually$, or equivalently, the projection Hamiltonian in the two ancilla qubits in each filter box, Fig.(\ref{filterbox}), does project appreciable part of state on the second last row. This assumption may not be correct in many SAT problems, especially close to $\alpha_c$.

In the circuit of Fig.(\ref{circuit}), if there is at least one  solution, and all electrons are simultaneously found on the final rows of all qubits, then the reading of the $N$-bit state satisfies all clauses.

In order to keep the energy gap from being too small, like in \cite{ours}, on every qubit teleportation boxes are inserted between two control Hamiltonians, thus the total number of qubits increases while the energy gap $\Delta\propto\epsilon/\lambda^8$ if all the boost and the projection Hamiltonians have the same value of amplifying factor $\lambda$.

 For one clause, or a filter box,  it needs 10 teleportation boxes (each teleportation box adds two more qubits) on the original five-qubit circuit, noting that  on the end of qubit $i,\ j$ and $k$ in Fig.(\ref{filterbox}) teleportation boxes are needed because more clause will be added. Thus adding one more filter box means adding 20 more qubits.
 The  number of clause for a NP hard 3-bit Exact Cover problem is about the same order as the number of bits $N$\cite{3SAT}, say $ \alpha N$ with $\alpha$ being $O(1)$, then there are about $20\alpha N$ qubits and each of them ends with either a projection or a boost Hamiltonian. Probability of finding all electrons at the final rows is approximately
\eqa
P\approx \left(1-{C}/{\lambda^2}\right)^{20\alpha N},
\eea
where $C=8$, the length of the longest qubit\cite{ours}. It is assumed that, at ground state, in each filter box the ancilla qubit 1 and 2 have appreciable probability in $|1\rangle$ and $|0\rangle$ states, respectively, before projection Hamiltonians. Later we will address the situation when this  assumption is violated.

In order to make the probability independent of number of bits $N$, we take $\lambda^2=DN$, where $D$ is an arbitrary number. Then as $N$ becomes large, we obtain
\eqa
P\approx\left(1-{C}/{(DN)}\right)^{20\alpha N}\approx e^{-20\alpha C/D},
\label{P}
\eea
and energy gap is\cite{ours}
\eqa
\Delta\propto \epsilon/\lambda^8\propto {\epsilon}/{(D^4N^4)},
\eea
from which one can estimate time cost.

To make the GSQC circuit at ground state,
 we can use adiabatic approach: first we set $\lambda=1$ for boost and projection Hamiltonian on final rows of all qubits, and replace the single qubit Hamiltonian between the first two rows of all qubits by a boost Hamiltonian 
\eqa
h'(B,\lambda')=\epsilon\left[
\frac{1}{\lambda'^2}  C^{\dagger}_{1}C_{1}+  C^{\dagger}_{2}C_{2}
-\frac{1}{\lambda'}\left(C^{\dagger}_{1}C_{2}+h.c.\right)\right],
\eea
so that the wavefunction amplitude of the first row is boosted as $\lambda'\gg 1$. Now in the ground state the electrons concentrate at the first rows as $1/\lambda'\rightarrow 0$, thus the ground state is easy to be prepared, and the energy gap $\Delta\propto\epsilon/n^2$ with $n= 8$ being the length of the longest qubit. The next step is turning the quantity $1/\lambda'$ to 1 adiabatically, during which the energy gap remains at $\epsilon/n^2$ and the ground-state wavefunction spreads to other rows from the first row.  The third step is turning $1/\lambda$ from 1 to $1/\sqrt{DN}$ adiabatically. In this process the energy gap   decreases monotonically from $\epsilon/n^2$ to what we obtained above: $\epsilon/D^4N^4$, and the ground-state wavefunction concentrates on the final rows of all qubit as we wish. Thus the scale of time cost is about $T\propto 1/\Delta^2\propto N^8$\cite{Farhi0}, local adiabatic approach may reduce the time cost further\cite{local}.

\subsection{Energy Gap for SAT Problems}

Above analysis is under the assumption that the number of satisfying states gradually decreases as the clauses are implemented one by one. There is a situation that might hurt our algorithm: 
after adding one more clause, the number of satisfying states drops dramatically. Just like what happens to Grover's search algorithm, in which the number of satisfying states drops from $2^N$ to 1, and as shown in Eq.(\ref{fff}), our algorithm involves a projection Hamiltonian on an ancilla qubit to project an exponentially small fraction of a state, thus the energy gap evaluation in the above subsection becomes invalid.

 Does this happen to the general SAT problems? In \cite{nature} it was suggested that close to the threshold $\alpha_c$   computational complexity might be related with the forming of a backbone, each of a subset of bits has average value close to 1 or 0 in the subspace of satisfying states. The existence of the backbone means that most satisfying states contain the state represented by the backbone, and if adding one more clause kicks out the states consistent with the backbone from satisfying subspace, the number of satisfying states drops dramatically, and this corresponds to projecting a small fraction of state. 

Performance of  our algorithm is not affected by forming of backbone, however, as more clauses applied, the disappearance of the already existed backbone in the satisfying subspace surely hurts.  There is a criterion determining efficiency of our algorithm: the ratio $S_j/S_{j+1}$, with $S_j$ being the number of solutions when the $j$th clause is applied, and $S_{j+1}$ the number of solutions when the $(j+1)$th clause is applied. For example, $S_0/S_1=8/3$ for 3-bit Exact Cover problem.
 If $S_j/S_{j+1}\gg 1$, on the ancilla qubit of the $(j+1)$th filter box, the probability of finding electron on its final row will be $p\approx(1-C S_j/(\lambda^2 S_{j+1}))$. To make sure of appreciable probability of finding all electrons on the final row of all qubits, an overhead factor $\sqrt{{S_j}/{S_{j+1}}}$ for $\lambda$ on the ancilla qubit is needed, hence the amplifying factor in the projection Hamiltonian on the ancilla qubit should be $\lambda \sqrt{{S_j}/{S_{j+1}}}$. 
According to the analysis in Sec.\ref{fraction}, the energy gap might be also determined by the parameter $S_j/S_{j+1}$. Because in a filter box, the ancilla qubit will end after the projection Hamiltonian, which should be at the position of qubit 8 or qubit 10 in Fig.(\ref{teleport}) without the dotted line following. According to the rule described in section \ref{fraction}, the parameter $1/x$ on this ancilla qubit should be
\eqa
\frac{1}{x}=\frac{1}{\lambda^2 Min\left(\lambda, \sqrt{\frac{S_j}{S_{j+1}}}\right)}.
\eea
 The energy gap thus is
\eqa
\Delta=Min\left(\frac{\epsilon }{\lambda^8},\frac{\epsilon S_{j+1}}{\lambda^4{S_j}}\right).
\eea
 If this ratio $S_{j+1}/S_{j}$ happens to be exponentially small, then our algorithm cannot solve the SAT problem in polynomial time. We cannot know in advance what $S_{j+1}/S_{j}$ is, however, we might be able to identify backbone by trials, and then choose proper order to implement clauses so that $S_{j+1}/S_{j}$ always can be kept not too small. However, if the  NP-Complete problem means that one can never avoid an exponentially small $S_{j+1}/S_j$, then the quantum algorithm cannot solve NP-Complete problem in polynomial time.

\section{Conclusion}

In conclusion, we have demonstrated that a ground state quantum computer can solve a general SAT problem. A specific example, the 3-bit Exact Cover problem, is given.
We show that a 3-bit Exact Cover problem can be solved by the quantum algorithm described here, and  the time cost is related with the number of bits $N$ and the parameter $S_{j+1}/S_{j}$. If $S_{j+1}/S_{j}$ stays only polynomially small, then the presented algorithm can solve this SAT problem in polynomial time. It will be interesting if one finds the equivalent algorithm by standard paradigm. 

I would like to thank A. Mizel for helpful discussion.  This
work is supported in part by Jun Li Foundation, the NSF under grant \# 0121428, and
ARDA and DOD under the DURINT grant \# F49620-01-1-0439.

\end{document}